# Multi-Technique Characterization of Rhodium Gem-Dicarbonyls on TiO$_2$(110)

Moritz Eder,[1] Faith J. Lewis,[1] Johanna I. Hütner,[1] Panukorn Sombut,[1] Maosheng Hao,[2] David Rath,[1] Jan Balajka,[1] Margareta Wagner,[1] Matthias Meier,[3] Cesare Franchini,[3,4] Ulrike Diebold,[1] Michael Schmid,[1] Florian Libisch,[2] Jiri Pavelec*,[1] and Gareth S. Parkinson[1]

[1] Institute of Applied Physics, TU Wien, Vienna, Austria

[2] Institute of Theoretical Physics, TU Wien, Vienna, Austria

[3] Faculty of Physics and Center for Computational Materials Science, University of Vienna, Vienna AT 1090, Austria

[4] Dipartimento di Fisica e Astronomia, Università di Bologna, Bologna IT 40126, Italy

*Corresponding authors: pavelec@iap.tuwien.ac.at

**Abstract**

Gem-dicarbonyls of transition metals supported on metal (oxide) surfaces are common intermediates in heterogeneous catalysis. While infrared (IR) spectroscopy is a standard tool for detecting these species on applied catalysts, the ill-defined crystallographic environment of species observed on powder catalysts renders data interpretation challenging. In this work, we apply a multi-technique surface science approach to investigate rhodium gem-dicarbonyls on a single-crystalline rutile TiO$_2$(110) surface. We combine spectroscopy, scanning probe microscopy, and Density Functional Theory (DFT) to determine their location and coordination on the surface. IR spectroscopy shows the successful creation of gem-dicarbonyls on a titania single crystal by exposing deposited Rh atoms to CO gas, followed by annealing to 200–250 K. Low-temperature scanning tunneling microscopy (STM) and non-contact atomic force microscopy (nc-AFM) data reveal that these complexes are mostly aligned along the [001] crystallographic direction, corroborating theoretical predictions. Notably, x-ray photoelectron spectroscopy (XPS) data reveal multiple rhodium species on the surface, even when the IR spectra show only the signature of rhodium gem-dicarbonyls. As such, our results highlight the complex behavior of carbonyls on metal oxide surfaces, and demonstrate the necessity of multi-technique approaches for the adequate characterization of single-atom catalysts.

**1. Introduction**

Transition metal carbonyl compounds, including those of Rh, Pd, and Pt, play a pivotal role in both homogeneous and heterogeneous catalysis.[1] This versatility arises from the strength of the metal–CO bond, which is strong enough to stabilize the compound yet weak enough to render CO an exchangeable ligand. As a result, transition metal carbonyls are established catalysts for a number of processes, such as acetic acid synthesis, hydroformylation, Reppe chemistry, and polymerization reactions.[2] Metal carbonyls are also integral to the field of single-atom catalysis (SAC). An isolated single atom is typically prone to sintering when they lack sufficient coordination on the surface, but such species can be stabilized by coordinating to suitable ligands. Transition metal carbonyls are therefore a direct link between homogeneous and heterogeneous SAC, since the single atom anchored to the surface resembles a metal complex bound to (exchangeable) ligands.[3–6] Therefore, transition metal carbonyls are garnering significant attention as precursors or intermediates in SAC systems.[7]

For noble metals like Pt, Ir, Pd, and Rh, single-atom dicarbonyls are frequently observed on supports such as zeolites, alumina, ceria, and titania.[8,9,7] These species, often referred to as geminal (gem-)dicarbonyls, have been extensively studied, particularly in the case of Rh due to its relevance in e.g. catalytic converters or the hydroformylation reaction.[10–12] IR spectroscopy has been the primary technique for detecting gem-dicarbonyls, which yield two characteristic peaks originating from the symmetric and asymmetric vibrational modes of the CO molecules. For Rh, these are typically located between 2120–2075 cm$^{-1}$ and 2053–1989 cm$^{-1}$, respectively.[9] Recently, Rh carbonyl species have once more received considerable attention as surface species involved in catalytic reactions.[13–19] Christopher and co-workers have expanded the scope by experimental and computational studies of Rh gem-dicarbonyls on various metal oxide supports for applied catalysis.[17,20,21] Their works elucidate the creation and stability of Rh carbonyls under various conditions. This includes a study of different Rh carbonyls on rutile $TiO_2$ that form after activation of the catalyst by heating in CO. Complementary DFT calculations by Sautet and co-workers shed light on their relative stability as a function the chemical potentials of $O_2$ and CO.[16] This particular work demonstrated how Rh adatoms adapt to the conditions dictated by the environment, and that the gem-dicarbonyl is the dominant form under reducing conditions in the presence of CO.

Despite these advancements in understanding transition metal carbonyls on oxidic supports, comprehensive atomic-level studies using diverse experimental techniques remain scarce, and IR spectroscopy remains the main tool for their investigation on powders. However, the IR spectral bands for gem-dicarbonyls in the literature span a range of approximately 50 cm$^{-1}$, and band intensities can be unreliable as quantitative measures.[9] The transition dipole moment of a molecule determines its signal intensity and can differ vastly depending on the adsorption site and the interplay of donation and back-donation of electron density. This effect is especially pronounced for the typical probe molecule CO, making the use of additional analytic techniques particularly important for characterizing carbonyl species.[7,9,22] Moreover, conventional DFT methods to determine the calculated frequencies of the IR absorption bands are often inaccurate.[23] To address these challenges, this study employs a multi-technique approach on an idealized model system. Using a rutile $TiO_2$(110) single crystal in ultra-high vacuum (UHV), we leverage a newly developed infrared reflection absorption spectroscopy (IRAS) setup,[24] linking our findings to those in the literature with IR spectroscopy as a bridging technique. Complementary low-temperature STM and nc-AFM provide direct visualization of Rh gem-dicarbonyls on the surface, pinpointing their location and coordination on $TiO_2$(110). Additionally, XPS data provide information about the chemical state of Rh on the surface. DFT calculations rationalize the experimental findings from IRAS and XPS. This comprehensive approach establishes a benchmark for Rh gem-dicarbonyls and underscores the potential pitfalls of relying solely on a single technique for characterizing (single-atom) surface species.

## 2. Experimental and Computational Methods

IRAS, XPS, and temperature-programmed desorption (TPD) experiments were conducted in a UHV system for surface reactivity studies[25] using a $TiO_2$(110) (5 × 5 × 0.5 $mm^3$) single crystal from CrysTec GmbH. The samples were mounted with Ta clips onto a Ta backplate, using a thin Au sheet in between to improve thermal contact. The sample was cooled using a liquid-He flow cryostat and heated to the desired temperature by resistively heating the Ta backplate. The vacuum chamber is equipped with a home-built effusive molecular beam source, where the gas streams out of an orifice with an effective diameter of 38.0 ± 1.9 µm. This source delivers an almost ideal top-hat profile at the sample with a 3.5 mm diameter and beam core pressure of 3.0 ± 0.3 × $10^{-8}$ mbar at the sample position.[26] The base pressure in the chamber was below $10^{-10}$ mbar. A quadrupole mass spectrometer (Hiden HAL 3F PIC) was used in a line-of-sight geometry for TPD experiments. Monochromatized Al $K_\alpha$ irradiation from an Al/Ag twin anode X-ray source (Specs XR50 M, FOCUS 500) and a hemispherical analyzer (Specs Phoibos 150 with a delayline detector) were used for XPS measurements. The energy scale is calibrated after each bakeout using copper, silver, and gold foils attached to the cryostat. The IRAS spectra were recorded using our newly developed setup for measuring dielectrics.[24] The [001] surface vector of the $TiO_2$(110) sample is in the plane of incidence. This implies that for p-polarized light, the electric field component parallel to the surface-parallel oscillates predominantly in the direction of the Ti and O rows, and that of s-polarized light is perpendicular to the rows. The p signal is more pronounced in our design compared to other setups due to the selected non-grazing angular range for p-polarized light on TiO$_2$(110). The range of incidence angles was set to 48°–65° for p-polarized light (i.e., the non-grazing side of the Brewster angle). The IRAS spectra in this work originate from the difference of the sample reflectivity spectrum $R$ that of the adsorbate-free surface, $R_0$. The normalized reflectivity difference is defined as

$$\frac{\Delta R}{R_0} = \frac{R-R_0}{R_0}.$$

The sample was cleaned by cycles of sputtering (15 min, 1 keV Ne$^+$, $I_{sample} \approx$ 1 µA/cm²) and annealing (900 K, 15 min). As a final step, the sample was first oxidized by annealing in O$_2$ (5 × 10$^{-7}$ mbar O$_2$, 900 K, 30 min) to avoid overreduction, and then vacuum-annealed (900 K, 15 min).[27,28] XPS scans after the cleaning procedure confirmed the absence of impurities on the surface within the detection limit. The density of oxygen vacancies was ≈12% with respect to the number of surface unit cells, as judged from D$_2$O TPD spectra. Rh was deposited using a water-cooled e-beam evaporator (FOCUS EFM3), whose flux was calibrated using a temperature-stabilized quartz microbalance (QCM). One monolayer (ML) corresponds to 1 Rh atom per surface unit cell.

Scanning probe microscopy (SPM) was done in a separate UHV system consisting of an analysis chamber with a base pressure lower than 10$^{-11}$ mbar and a preparation chamber with a base pressure lower than 10$^{-10}$ mbar. The sample was prepared in the same way as for the XPS and IRAS measurements, with annealing temperatures of ≈1100 K. The O defect density amounted to ≈11% of all surface unit cells as determined by STM. The images were obtained using a qPlus sensor[29] ($f_0$ = 31.8 kHz, $k$ = 1800 N/m, $Q$ ≈ 10000) with an electrochemically etched tungsten tip. Typical oscillation amplitudes were $A$ = 150 pm. The base temperature in the SPM head was 4.7 K, but the sample was counter-heated to 14 K in order to provide sufficient conductivity for STM, unless otherwise noted. STM was measured with a positive sample bias, tunneling into the empty states of the sample ($U$ = +0.5 V to +3.0 V). An atomically sharp metal tip was prepared on a Cu(110) single crystal by voltage pulses and dipping the tip into the surface. Images were processed by correcting for the piezo creep and drift[30] and filtering few frequencies of mechanical noise in the Fourier domain. Rh was deposited using an e-beam evaporator (EFM3, FOCUS) cooled with liquid nitrogen, with the flux calibrated using a temperature-stabilized QCM, and 1 ML corresponding to 1 Rh atom per surface unit cell.

*Computational methods*

All calculations were performed using the Vienna *ab initio* simulation package (VASP).[31] The projector augmented wave (PAW) method[32,33] was employed for the near-core regions and the plane-wave basis set cutoff energy was set to 700 eV. To more accurately capture the electronic structure around the Rh atom, we used 17 active electrons, and accordingly the VASP PAW potentials O_h, C_h, Rh_sv_GW and Ti_sv. Calculations were performed using optPBE-vdW, a spin-polarized GGA method. This approach integrates the optPBE exchange-correlation functional[34] with non-local correlation corrections from vdW-DF as proposed by Dion et al., effectively incorporating van der Waals interactions.[35,36] An effective on-site Coulomb repulsion term $U_{eff}$ = 3.9 eV was applied to the *d*-orbitals of the Ti atoms.[37] The unreconstructed rutile TiO2(110) surface was modeled as an asymmetric slab comprising five $TiO_2$ tri-layers within a large two-dimensional 6 × 2 unit cell and including a vacuum region greater than 12 Å along the z-axis. The top three tri-layers were allowed to relax, while the bottom two tri-layers were kept fixed at their bulk positions. Additionally, pseudo-hydrogen atoms were used to saturate the bottom layer to fulfill the octet rule; this makes the bottom surface more bulk-like and thereby allows using thinner slabs.[38] The convergence criterion was an electronic energy step of $10^{-7}$ eV and forces acting on ions smaller than 0.01 eV/Å.

The adsorption energies were computed according to the formula

$$E_{ads} = \left(E_{Rh/TiO_2+nCO} - \left(E_{Rh/TiO_2} + nE_{CO}\right)\right)/n$$

where $E_{Rh/TiO_2+nCO}$ is the total energy of the Rh adatom on TiO2(110) surface with n adsorbed CO molecules, $E_{Rh/TiO_2}$ is the total energy of the adsorbed Rh adatom on the TiO2(110) surface, and $E_{CO}$ represents the energy of the CO molecule in the gas phase. The Rh 3d core-level binding energies were calculated using the initial state approximation.[39,40] Atomic charges were determined through Bader charge analysis.[41] The CO vibrational frequencies were computed within the harmonic approximation, employing the finite-difference method.

For improved accuracy in frequency calculations, we used the HSE06 hybrid functional[42] with the standard mixing factor 25% and a screening length of $0.2^{-1}$ Å$^{-1}$. The computed CO stretching vibrational frequency for adsorbed CO on the Rh/TiO2(110) was scaled by the method-dependent factor $v_{CO_{gas}}^{exp}/v_{CO_{gas}}^{cal}$, with $v_{CO_{gas}}^{exp}$ = 2143 cm$^{-1}$, and $v_{CO_{gas}}^{cal}$ = 2114 cm$^{-1}$ and 2232 cm$^{-1}$ for optPBE-vdW and HSE06, respectively. The geometry was fully reoptimized accordingly, and all HSE06 calculations were performed using a plane-wave basis set cutoff energy of 400 eV with standard PBE pseudopotentials. The convergence criteria were set to an electronic energy threshold of $10^{-6}$ eV and ionic forces below 0.01 eV/Å.

## 3. Results

### 3.1 IRAS Spectra of Rh Gem-Dicarbonyls

#### 3.1.1 Experimental IRAS Frequencies

Two methods for synthesizing Rh gem-dicarbonyls on $TiO_2$(110) in UHV are reported in the literature, either by decomposing Rh clusters under high pressures of CO gas,[43] or by decomposing $Rh(CO)_2Cl_2$ complexes.[44] However, the first approach was not possible in a clean fashion with our experimental setups, and the latter yields significant amounts of Cl on the surface, which we sought to avoid due to its potential impact on the electronic and geometric structures of the system.[45] We have shown previously that single Rh atoms can be immobilized on $TiO_2$(110) by evaporating Rh onto the surface at 80 K,[46] and we used this method as a basis to synthesize the Rh gem-dicarbonyls as illustrated in the following. **Figure 1** shows IRAS spectra of 0.05 ML Rh on $TiO_2$(110) deposited at 80 K before and after CO adsorption. The spectra were recorded sequentially at 80 K. The accompanying annotations describe the specific sample treatments applied prior to each measurement.

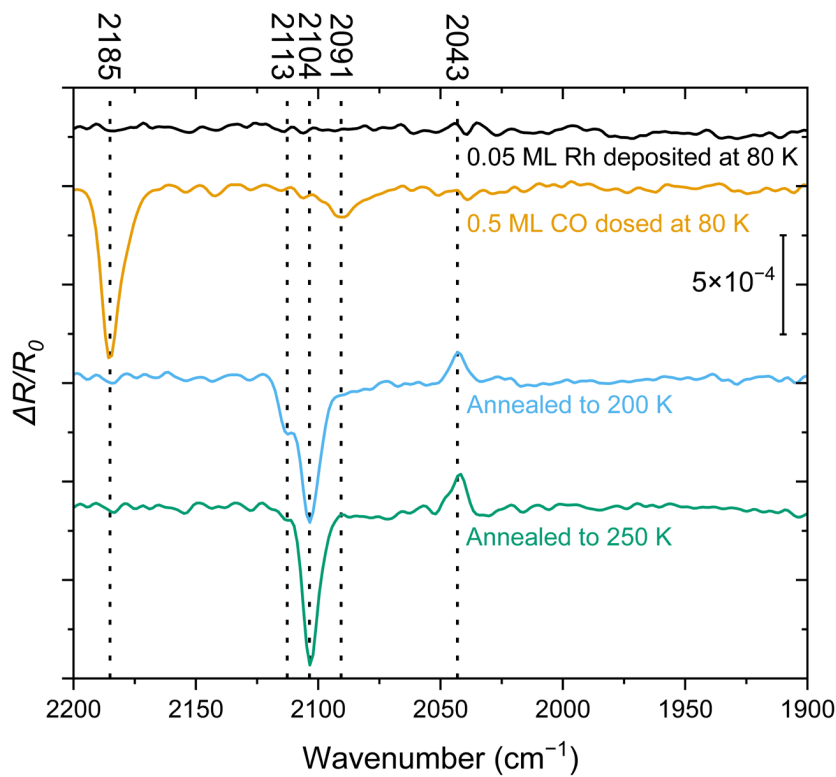

**Figure 1**. IRAS spectra (baseline-corrected, obtained with p-polarized light, 4 cm$^{-1}$ resolution, 4000 scans, 60 kHz mirror velocity, each recorded at 80 K over ≈10 min) of 0.05 monolayers (ML) Rh deposited at 80 K on TiO$_2$(110) (black trace), with additional 0.5 ML CO dosed at 80 K (orange trace), after heating to 200 K (blue trace), and after heating to 250 K (green trace). The peaks at 2104 cm$^{-1}$ and 2043 cm$^{-1}$ are assigned to the symmetric and asymmetric vibrational modes of the Rh gem-dicarbonyl, respectively. The reference spectrum was recorded before the series and used for all subsequent measurements.

The spectrum recorded immediately after Rh deposition at 80 K (**Figure 1**, black) does not show any detectable features in the relevant wavenumber range of 2200–1900 cm$^{-1}$ where the CO-stretch vibration of transition metal carbonyls is commonly found. This indicates that CO adsorption from the residual gas in the UHV chamber is negligible. After adding 0.5 ML CO at 80 K (**Figure 1**, orange), we identify two signals: an intense peak at 2185 cm$^{-1}$, attributed to CO on TiO$_2$(110),[24,47] and a band around 2091 cm$^{-1}$. The latter falls within a region where multiple species of Rh carbonyls have been postulated on metal oxides and zeolites.[9,48,49] The rather broad peak suggests a mixture of species. After flashing the sample to 200 K and cooling it back to 80 K (**Figure 1**, blue), the CO/TiO$_2$(110) peak has disappeared, consistent with the desorption of CO from this surface at ≈150 K.[47] Furthermore, the feature at 2091 cm$^{-1}$ is no longer present. Instead, we observe three species at 2113 cm$^{-1}$, 2104 cm$^{-1}$, and 2043 cm$^{-1}$. The latter two are characteristic of the Rh gem-dicarbonyl[44] and align with the expected parallel and perpendicular contributions of p-polarized light over the measured angular range.[24] The azimuthal orientation of the gem-dicarbonyls on single-crystalline metal oxides can therefore be inferred from the IR data, since the surface selection rule that governs reflectivity on metallic surfaces does not apply on dielectric substrates.[24,44,47,50,51] In our setup, the principal ray of the IR beam projected onto the surface is parallel [001], i.e., parallel to the Ti and O rows. The azimuthal orientation of the gem-dicarbonyls on single-crystalline metal oxides can be inferred from the appearance of the positive Δ$R$/$R_0$ peak; this peak would not appear in p-polarization for asymmetric stretch in the [1$\bar{1}$0] direction (an azimuthal orientation of the gem-dicarbonyl perpendicular to the Ti and O rows). When Rh dicarbonyls are measured on powders whose surfaces are randomly oriented, these two peaks are observed as two minima of the reflectance (maxima of absorbance) in IR spectra. However, our IR beam is reflected on a single crystal surface at a well-defined range of incidence angles, chosen such that the symmetric (2104 cm$^{-1}$) and asymmetric stretch (2043 cm$^{-1}$) result in a lower and higher reflectivity, respectively. This results from the orientation of the dipole moments and Fresnel equations, as confirmed by simulations.[24] Following annealing to 250 K (**Figure 1**, green trace) the 2113 cm$^{-1}$ feature disappears, while the other two signals associated with the gem-dicarbonyl remain. We cannot assign the peak at 2113 cm$^{-1}$ to one certain species at this point, but we consider it to be a metastable Rh-carbonyl species.[9,48] By annealing to room temperature and above (**Figure S1**), The Rh dicarbonyls disappear and a band around 2020–2040 cm$^{-1}$ emerges, where CO on metallic Rh is typically located.[52] This is clear evidence that the previously observed Rh species have agglomerated to form CO-covered Rh clusters. While we have observed these vibrational frequencies of the carbonyl species most frequently in our experiments, we have occasionally observed a slight shift of 1–4 cm$^{-1}$ towards higher wavenumbers. We ascribe this to differences in the preparation procedure, and to an enhanced number of adsorbates from the residual gas of the UHV chamber, interacting with the gem-dicarbonyls when the adsorbates are in close proximity to them.

An alternative synthetic route towards the Rh gem-dicarbonyls involves depositing Rh onto a CO-covered TiO$_2$(110) surface. This approach has been successfully employed for creating Fe complexes on ligand-covered FeO(111) single crystals.[53,54] **Figure S2** shows the corresponding IRAS spectra. This method results in the formation of multiple Rh carbonyl species, even after annealing. It hence appears less effective for the selective synthesis of a single type of Rh gem-dicarbonyls on TiO$_2$(110).

### 3.1.2 DFT Calculations on Rh Gem-Dicarbonyls

The optimized ground-state structure of the Rh gem-dicarbonyl was determined using DFT+U calculations, (**Figure 2 c**) illustrates its arrangement on the TiO$_2$(110) surface: The Rh atom (light grey) is positioned between two bridge-bonded O anions (red). The two CO molecules (with C in black) attached to the Rh atom are aligned parallel to the [001] direction. This is the lowest-energy Rh gem-dicarbonyl configuration found in our calculations and an earlier work by Sautet and co-workers.[16] The adsorption energy per CO molecule in the square-planar geometry is energetically more favorable by nearly 0.65 eV compared to the tetrahedral alternative, in which the CO molecules are aligned perpendicular to the O and Ti rows (**Figure S3**). The calculated Bader charge of the Rh in the gem-dicarbonyl of **Figure 2 c** is +0.68 e, which is typical for a $Rh^{1+}$ state. This means that one Ti atom in the subsurface layer is reduced from $Ti^{4+}$ to $Ti^{3+}$.[46,55] These findings align with the well-known preference of $Rh^+$ complexes for square-planar over tetrahedral geometries, as observed in various coordination compounds.[56]

Our calculated CO vibrational frequencies are 2098.9 $cm^{-1}$ and 2042.6 $cm^{-1}$ at the HSE06 level (2080 $cm^{-1}$ and 2021 $cm^{-1}$ with the optPBE-vdW functional). These values align well with our experimental observations within the accuracy of DFT, particularly regarding the difference between symmetric and asymmetric stretching modes (61 $cm^{-1}$ in the experiment and 59 $cm^{-1}$ in the HSE06 calculations). The difference between the two modes exhibits a smaller error than their individual absolute frequency values due to error compensation from the shared approximation. We considered the possibility of Rh monocarbonyl species using optPBE-vdW, but our calculated CO vibrational frequency with optPBE-vdW for a Rh monocarbonyl as shown in **Figure S3** is 2011 $cm^{-1}$, which is not observed in our IR results. Therefore, we ascribe the IR peaks at 2104 $cm^{-1}$ and 2043 $cm^{-1}$ to the symmetric and asymmetric stretch of $Rh^+(CO)_2$, respectively.

## 3.2 Rh Gem-Dicarbonyls in Low-temperature Scanning Probe Microscopy

To further confirm the conformation of Rh gem-dicarbonyls, we employed low-temperature STM and nc-AFM. **Figure 2** presents STM and nc-AFM images taken at a sample temperature of 14 K. The sample was prepared by depositing 0.05 ML Rh onto the $TiO_2$(110) surface, followed by dosing 1 Langmuir (L) of CO and annealing to ≈270 K. An overview scan is presented in **Figure S4**.

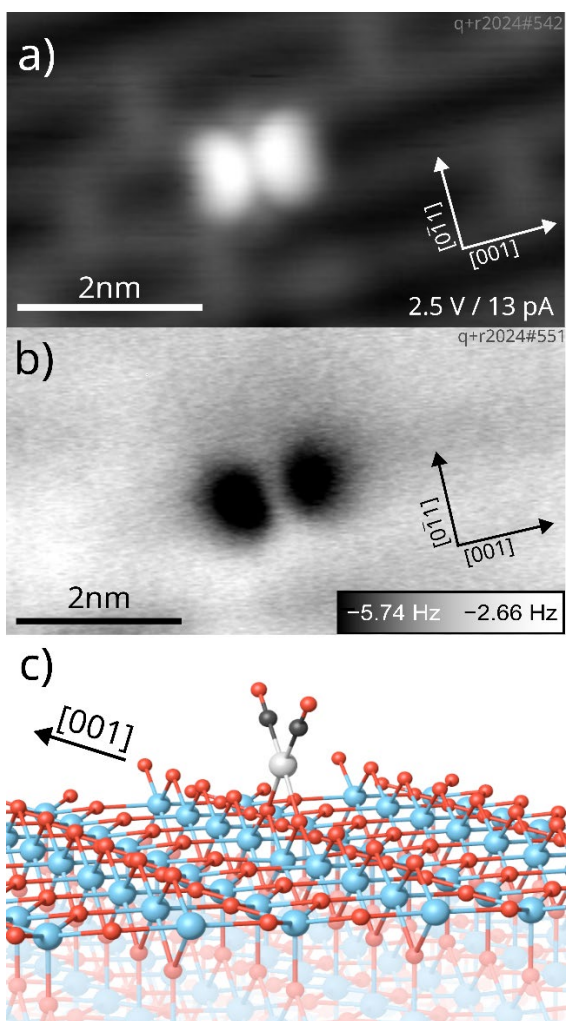

**Figure 2.** Low-temperature scanning probe images of a Rh gem-dicarbonyl on $TiO_2$(110) taken at 14 K with a Cu-terminated tip. a) Empty-state STM image showing the two CO molecules of the Rh

gem-dicarbonyl appearing as bright lobes on the TiO$_2$(110) surface. b) Constant-height nc-AFM image of the same gem-dicarbonyl with a metallic tip, imaging the two CO molecules as dark circles due to the attractive interaction between the metal tip and the CO (amplitude = 150 pm, $U$ = 0.0 V). c) The optimized structure of the Rh gem-dicarbonyl on the TiO$_2$(110) surface obtained by DFT+U calculations (Ti: blue; O: red; C: black; Rh: grey).

The overview scan in **Figure S4** shows the presence of clusters apart from the features we ascribe to the gem-dicarbonyls (green circles). We believe that the abundance of clusters is due to the fact that the annealing temperature in the STM experiment (≈270 K) was higher than in the IR experiment (250 K). However, it is plausible that clusters are present on the sample used in the IR experiment as well, but are not detected if they have no CO molecules adsorbed. In empty-state STM images in **Figure 2a**, we observe the characteristic features of the TiO$_2$(110) surface.[27] The bright and dark rows along the [001] direction correspond to the Ti cations and bridge-bonded O anions, respectively. Bright spots on the black rows are O vacancies,[57] evidence for a reduced, non-stoichiometric surface. We find a distinct feature in the form of a bright double-lobe oriented along the [001] direction, located on top of a bridge-bonded O row. The center of the lobes is exactly located between two bridging oxygen anions, as illustrated by the unit cell grid centered on the oxygen vacancies (see **Figure S5**). The bright double lobe stems from the two CO molecules aligned in a plane parallel to the Ti and O rows, in excellent agreement with our spectroscopic and computational results, and calculations by Sautet and co-workers.[16] A nc-AFM image of the same spot on the surface in **Figure 2b** corroborates this interpretation. We find again two lobes aligned in the [001] direction, which appear dark in this case due to the attractive interaction between the metallic Cu-terminated tip and the oxygen of the CO molecules.

As mentioned above, the STM images show a substantial number of species attributed to $Rh_n$ clusters at a Rh coverage of 0.05 ML and successive CO adsorption and annealing to ≈270 K. By using the same preparation of Rh gem-dicarbonyls on the $TiO_2$(110) surface but aiming at a lower Rh coverage (0.005 ML, i.e., one tenth of the previous coverage), the number of large clusters is reduced, but Rh clusters cannot be fully avoided. **Figure 3** shows SPM images of the surface after this preparation.

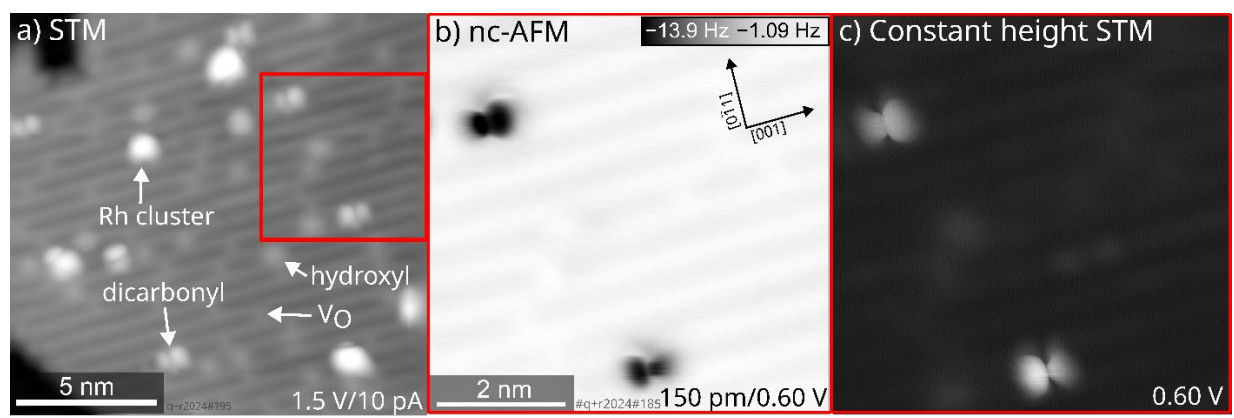

**Figure 3**. Low-temperature SPM images of a Rh gem-dicarbonyl on $TiO_2$(110) taken at 14 K with a Cu-terminated tip. 0.005 ML Rh were deposited at 100 K, exposed to 1 Langmuir (L) CO at 100 K, and heated to 270 K. a) Empty-state STM image of a 15 nm x 15 nm area. The red square represents the area of images b) and c). b) Constant-height nc-AFM image of two gem-dicarbonyls with a metallic tip, imaging the two CO molecules as black spots due to their attractive interaction to the tip (Amplitude = 150 pm). c) STM image of the same area as for b). The current was measured in constant-height mode during the nc-AFM scan.

An overview scan covering a larger surface area (**Figure 3 a)**) revealed Rh clusters as well as double-lobed, inequivalent features apart from the typical surface species like hydroxyls and O vacancies. A constant-height image of the red square in **Figure 3 a)** shows an asymmetry of two gem-dicarbonyls, depicted in the frequency shift in **Figure 3 b)** and the simultaneously recorded tunneling current **Figure 3 c)**. Under all imaging conditions, this asymmetry of the individual double-lobed features stays consistent. The opposite orientations of these inequivalent CO molecules suggest that the effect is not an artifact from a slightly asymmetric tip. Instead, the inequivalence could arise from distinct interactions between the two CO molecules of the dicarbonyl and the $TiO_2$(110) surface, causing a tilt of the structure in the [001] or [00$\bar{1}$] direction. One plausible explanation involves interaction with nearby OH groups, which are ubiquitous on $TiO_2$(110) surfaces. Our DFT calculations show that the relaxed interaction between one Rh-bound CO and a hydroxyl group on the O rows would indeed lead to an asymmetric position of the CO molecules with respect to the Rh (**Figure S6**). Recent calculations on anatase $TiO_2$(001) by Christopher and Pacchioni lend support to this hypothesis, demonstrating that OH groups can interact with CO molecules, influencing their geometry and alignment.[17] We therefore assign the different size of the lobes in nc-AFM to the coordination of a neighboring OH group with one CO molecule of the Rh gem-dicarbonyl complex.

### 3.3 XPS Spectra of Rh and C

To understand the chemical nature of the Rh species, we recorded XPS spectra during the same experiment as for the IRAS measurements depicted in **Figure 1**. **Figure 4** presents XPS spectra of the Rh $3d$ and C $1s$ region of the $TiO_2$(110) surface at 80 K after various treatments, as indicated in the corresponding annotations.

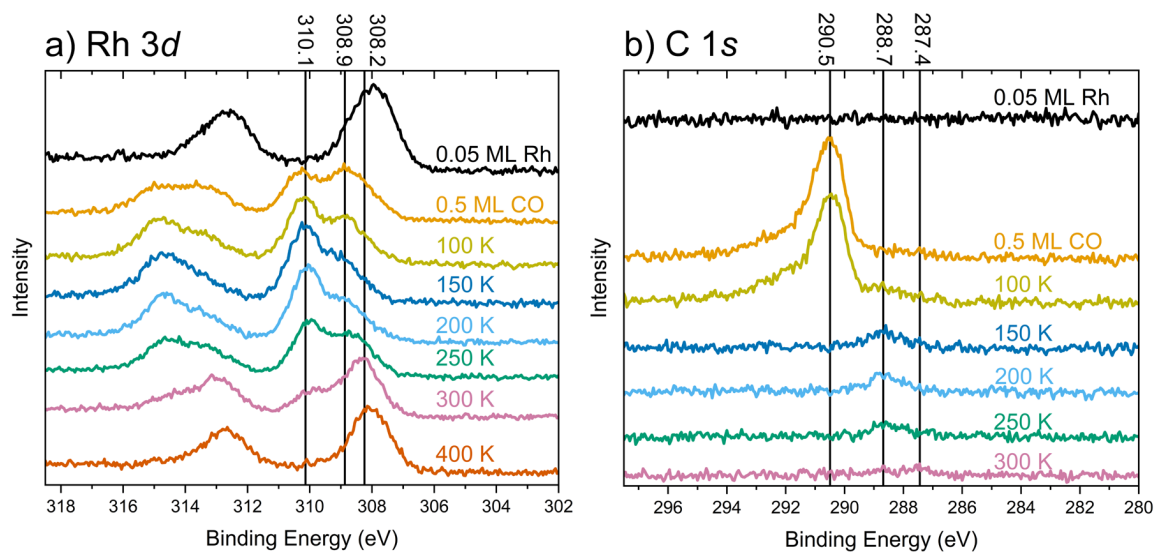

**Figure 4**. XPS spectra of a) the Rh 3*d* and b) C 1*s* regions taken at 80 K using monochromatized Al K$_\alpha$ radiation at 70° grazing emission for enhanced surface sensitivity. The colored text corresponding to each spectrum and indicates the treatment before the measurement.

The Rh 3*d* spectrum in **Figure 4a)** shows the expected doublet after depositing 0.05 ML Rh at 80 K (black trace). The Rh 3$d_{5/2}$ peak is centered at 308.1 eV, consistent with a prior study of this system.[46] The width of the peak suggests the coexistence of different species, presumably the single atoms and very small clusters as observed by STM/nc-AFM in **Figure S7**. Due to the broadness of the peak, the presence of metallic Rh, centered around 307.7 eV (**Figure S8**) seems likely. Adding 0.5 ML CO at 80 K (**Figure 4a)**, orange trace), splits the Rh 3$d_{5/2}$ peak into two maxima of comparable intensity. However, the peak broadness suggests contributions from different chemical surroundings. Upon heating to 100 K, 150 K, and 200 K (yellow, dark blue, and light blue traces, respectively), we still observe the two species at the same position, although their relative intensities change. At these temperatures, the higher-binding-energy component dominates. At 250 K (green trace), the intensity shifts slightly in favor of the lower-binding-energy component, and a shoulder emerges at 308.2 eV, indicative of a larger, agglomerated species. The observation of the component around 308.2 eV is consistent with the observation of clusters in STM in **Figure S4**. After annealing to 300 K (**Figure 4a)**, pink trace), the species at 308.2 eV dominates, with diminished intensities for the two components at higher binding energy. The substantial reduction in the intensity of the peak at 310.1 eV coincides with the loss of the bands from the gem-dicarbonyl in IRAS. Thus, we assign the peak at 310.1 eV to the Rh gem-dicarbonyl. This observation aligns well with a study by Hayden et al.,[58] who observed the agglomeration of Rh gem-dicarbonyls on $TiO_2$(110) at room temperature; this was attributed to the impact of hydrogen (i.e., hydroxyl groups) on the surface. Our result also aligns with observations by Frederick et al. found essentially the same Rh $3d$ binding energy (310.2 eV) for Rh gem-dicarbonyls on alumina.[59] The variety of Rh species in **Figure 4a)** is much larger after annealing to 300 K compared to depositing Rh directly at 300 K (see **Figure S8**). XPS of the latter yields only one significant Rh $3d_{5/2}$ peak at a neutral Rh position of 307.7 eV, originating from clusters as detected in STM.[46] After heating to 400 K (red in **Figure 4 a)**), we find one pronounced species around 308.1 eV, close to the initial position after Rh deposition in the absence of CO (307.9 eV). This peak is assigned to Rh clusters with a more metallic character. DFT core-level calculations using the initial state approximation reveal that the binding energies of Rh $3d$ for a Rh monocarbonyl and a Rh gem-dicarbonyl are shifted to higher values by 0.79 and 1.46 eV, respectively, relative to the bare Rh single-atomic ground-state configuration. A Rh $3d_{5/2}$ peak position around 310 eV could be taken as representative of a $Rh^{3+}$ charge state. However, the Bader charge analysis and the number of reduced Ti atoms indicate that both Rh species in the XPS spectra, with and without CO molecules attached, are in the Rh$^+$ state. Hence, the shift of the Rh peaks towards higher binding energies upon CO exposure is due to intramolecular redistribution of electron density, but not due to a change in formal charge state. This is in line with other single-atomic carbonyls on Fe$_3$O$_4$(001).[60] Our DFT results (**Figure S9**) show that increasing the Rh coordination by binding to the CO molecules in the square-planar configuration shifts the energies of d-band states to lower values compared to adsorbed, bare Rh. This also suggests that the available mobile charge around the Fermi edge to screen out the core hole is substantially reduced, which could lead to final state effects.

The C 1*s* spectra in **Figure 4b** exhibit no detectable signal for the Rh-decorated surface (black trace). Following CO dosing (orange trace), a peak emerges at 290.5 eV, which we ascribe to CO adsorbed on TiO$_2$. Heating to 100 K (yellow trace) induces a shoulder at 288.7 eV, which we ascribe to CO on Rh. This corroborates evidence from IRAS (**Figure 1**, orange trace) which suggests that CO is immobile after landing on the surface at 80 K and requires thermal energy to diffuse and to form Rh gem-dicarbonyls. After annealing to 150 K (dark blue trace), the peak of CO on TiO$_2$ (290.5 eV) has vanished due to CO desorption. This agrees again with the corresponding result in the IR spectra in **Figure 1** (blue trace) and the literature.[47] When increasing the temperature to 200 K and 250 K (**Figure 4b**) blue and green trace, respectively), the CO/Rh peak at 288.7 eV remains stable in position and intensity. At 300 K, we find its intensity diminished, due to the desorption of the CO from the Rh gem-dicarbonyls. This coincides with the emergence of a weak peak around 287.4 eV. This new signal is attributed to CO adsorbed on Rh clusters, reflecting the aggregation of Rh species as observed in the Rh 3*d* spectra.

## 4. Discussion

The infrared spectra (**Figure 1**) demonstrate the facile synthesis of Rh gem-dicarbonyls on $TiO_2$(110) in UHV, a model system for metal carbonyls anchored to metal oxides.[9,61] Our synthetic approach avoids high pressures and Cl contaminations on the surface (in contrast to previous works),[44,62] and is presumably transferable to other transition metals like Ir, Ni, Pt, or Pd, known for forming carbonyl complexes on metal oxide surfaces.[8,9,61] Our study provides the first low-temperature scanning probe images of Rh gem-dicarbonyls on $TiO_2$(110) (**Figure 2**). The revealed double-lobed features, aligned along the [001] direction, strongly support a square-planar geometry. This finding contrasts with earlier studies by Hayden et al., which provide evidence for a perpendicular alignment of the Rh gem-dicarbonyls to the Ti and O rows on $TiO_2$(110).[50] However, our spectroscopic results in **Figure 1** are consistent with our computational calculations and recent work by Tang et al., which predict the alignment along the bridging O rows on $TiO_2$(110),[16] as well as our SPM images. The square-planar geometry arises from the coordination of Rh with two CO ligands parallel to the Ti and O rows, reflecting the intrinsic preference of $Rh^+$ complexes for square-planar over tetrahedral conformations.[56] We have observed distortions of the dicarbonyls (**Figure 3**), possibly due to the interaction of CO ligands with surface OH groups, as suggested by DFT calculations (**Figure S6**). These distortions could influence the catalytic activity and selectivity of gem-dicarbonyl species.[63]

Our XPS data (**Figure 4**) provide critical insights into the electronic states of Rh species. Note that the Rh 3*d* peaks after deposition are located hat higher binding energies (≈308 eV for Rh $3d_{5/2}$) compared to neutral, bulk-like Rh (≈307 eV),[64] likely due to final-state effects of the small clusters.[46] The broadness of the signal is in perfect agreement with the fact that we find single atoms as well as small clusters in STM after deposition. Exposing the Rh surface species to CO and heating this system creates Rh carbonyls from single atoms and possibly the partial break-up of clusters.[65] The Rh 3*d* signal splits into a lower and higher BE component, the latter at 310.1 eV being dominant until 200–250 K (**Figure 4a**) and identified as originating from the gem-dicarbonyl, in agreement with the literature.[59] The authors of this work on alumina ascribed this peak to a $Rh^{3+}$ species, but our DFT results suggest that the Rh in the gem-dicarbonyl on $TiO_2$(110) is in the oxidation state 1+. We also rule out the oxidation state 2+ for the Rh species due the absence of multiplet splitting in the Rh 3*d* spectra.[66,67] The different Rh 3*d* binding energies with and without CO are rather due to electronic effects attributed to the different coordination (**Figure S9**). The calculated vibration frequencies confirm that the two main peaks observed in the same temperature range as the dicarbonyl signal in XPS (**Figure 1**, 2104 $cm^{-1}$ and 2043 $cm^{-1}$) are due to the dicarbonyl. The IRAS also shows a weak species at 2113 $cm^{-1}$, which disappears upon annealing to 250 K. We note that the disappearance coincides with a slight intensity increase of the gem-dicarbonyl peak at 2104 $cm^{-1}$, but also with an increase of the Rh 3*d* XPS signal at ≈308 eV (**Figure 4a**), which is related to small Rh clusters. Thus, it is unclear whether the species responsible for the 2113 $cm^{-1}$ peak transforms into the "normal" gem-dicarbonyl or creates Rh clusters upon its disappearance. In any event, upon annealing to temperatures above 250 K, the dicarbonyls vanish and the Rh forms small clusters. These observations are in line with studies proposing the reductive desorption of CO from Rh carbonyls on zeolites,[68] and the presence of clusters in the STM images in **Figure 3** and **Figure S4**.

These results show that relying solely on IR spectroscopy for identifying surface species, for example in single-atom catalysis, can be deceiving. More Rh species can be found in the XPS than one would expect from the IR data showing only a single dicarbonyl species following the 250 K anneal. This is likely due to these IR-invisible species binding to very few or no CO molecules, or the dipole moment of those CO species being too low.[22] Integration of the C 1$s$ peaks in **Figure 4b** shows that after CO desorption from the oxide (150 K and higher), the area under the remaining CO signal amounts to about 10% of the initial signal after dosing 0.5 ML CO. However, if all Rh atoms (0.05 ML) were covered by 2 CO molecules, it should amount to 20%. This provides additional evidence that not all Rh species are gem-dicarbonyls, and that some Rh atoms have to be in clusters, as suggested above, based on the lower binding energy component of the Rh 3$d$ signal.

The fact that the IR does not detect these significant quantities of clustered species has profound implications for catalysis, especially in single-atom systems, where the catalytic role of single atoms versus minor concentrations of clusters is often debated. A multi-technique approach, as demonstrated here, is essential to resolve ambiguities and provide a comprehensive understanding of the active surface species.[69]

**Conclusions**

Our results build upon and expand previous findings on gem-dicarbonyls on metal oxide supports. While earlier studies have primarily relied on IR spectroscopy, our multi-technique approach provides a more nuanced characterization of these species. The direct visualization of gem-dicarbonyls via STM and nc-AFM bridges the gap between theoretical predictions and experimental evidence. Their stability under realistic operating conditions is limited, with decomposition of the dicarbonyls and formation of Rh clusters occurring at temperatures above 250 K. However, the formation of gem-dicarbonyls at low temperatures and distinct spectroscopic fingerprints make them important intermediates in catalytic processes, potentially influencing reaction pathways. Additionally, the suggested interaction with surface OH groups may provide a route for tuning their electronic and geometric properties, opening avenues for site-specific catalysis. The scanning probe images show mainly gem-dicarbonyls oriented along the [001] direction along the Ti and O rows, in perfect agreement with the IRAS data and DFT calculations predicting a fourfold-planar coordination of $Rh^+$. The XPS data shows that there is more than one Rh species on the surface, attributed to small Rh clusters, invisible in the IR spectra despite being present in considerable amounts and almost certainly covered by CO at these temperatures. This demonstrates the necessity of multi-technique approaches in the investigation of single atoms and carbonyls, and has therefore important implications for single-atom catalysis.

**Acknowledgments**

Funding from the European Research Council (ERC) under the European Union's Horizon 2020 research and innovation program (grant agreement no. 864628,

Consolidator Research Grant "E-SAC") is acknowledged. This work was supported by the European Research Council under the European Union's Horizon 2020 research and innovation program (grant agreement No. 883395, Advanced Research Grant "WatFun"). This research was funded in part by the Austrian Science Fund (FWF) 10.55776/F81 and the Cluster of Excellence MECS (10.55776/COE5). For open access purposes, the authors have applied a CC BY public copyright license to any author accepted manuscript version arising from this submission. Moritz Eder acknowledges funding by the Marie Skłodowska-Curie Actions (Project 101103731, SCI-PHI). Matthias Meier gratefully acknowledges financial support from the Austrian Science Fund (FWF) (through project number 10.55776/PAT2176923). The computational results have been achieved using the Vienna Scientific Cluster (VSC). The authors acknowledge TU Wien Bibliothek for the financial support through its Open Access Funding Programme.

**References**

(1) Dyson, P. J.; McIndoe, J. S. *Transition Metal Carbonyl Cluster Chemistry*; CRC Press, 2018. https://doi.org/10.1201/9781315273815.
(2) Masters, C. *Homogeneous Transition-Metal Catalysis*; Springer Netherlands: Dordrecht, 1980. https://doi.org/10.1007/978-94-009-5880-7.
(3) Chen, F.; Jiang, X.; Zhang, L.; Lang, R.; Qiao, B. Single-Atom Catalysis: Bridging the Homo- and Heterogeneous Catalysis. *Chinese Journal of Catalysis* **2018**, *39* (5), 893–898. https://doi.org/10.1016/S1872-2067(18)63047-5.
(4) Kraushofer, F.; Haager, L.; Eder, M.; Rafsanjani-Abbasi, A.; Jakub, Z.; Franceschi, G.; Riva, M.; Meier, M.; Schmid, M.; Diebold, U.; Parkinson, G. S. Single Rh Adatoms Stabilized on α-$Fe_2O_3$ (1$\bar{1}$02) by Coadsorbed Water. *ACS Energy Letters* **2022**, *7* (1), 375–380. https://doi.org/10.1021/acsenergylett.1c02405.
(5) Kraushofer, F.; Parkinson, G. S. Single-Atom Catalysis: Insights from Model Systems. *Chemical Reviews* **2022**, *122* (18), 14911–14939. https://doi.org/10.1021/acs.chemrev.2c00259.
(6) Parkinson, G. S. "Single-Atom" Catalysis: An Opportunity for Surface Science. *Surface Science* **2025**, *754*, 122687. https://doi.org/10.1016/j.susc.2024.122687.
(7) Kaiser, S. K.; Chen, Z.; Faust Akl, D.; Mitchell, S.; Pérez-Ramírez, J. Single-Atom Catalysts across the Periodic Table. *Chemical Reviews* **2020**, *120* (21), 11703–11809. https://doi.org/10.1021/acs.chemrev.0c00576.

(8) Frank, M.; Bäumer, M.; Kühnemuth, R.; Freund, H.-J. Metal Atoms and Particles on Oxide Supports: Probing Structure and Charge by Infrared Spectroscopy. *The Journal of Physical Chemistry B* **2001**, *105* (36), 8569–8576. https://doi.org/10.1021/jp010724c.

(9) Hadjiivanov, K. I.; Vayssilov, G. N. Characterization of Oxide Surfaces and Zeolites by Carbon Monoxide as an IR Probe Molecule; 2002; pp 307–511. https://doi.org/10.1016/S0360-0564(02)47008-3.

(10) Yang C., G. C. W. Infrared Studies of Carbon Monoxide Chemisorbed on Rhodium. *The journal of physical chemistry.* **1957**, *61* (11), 1504. https://doi.org/10.1021/j150557a013.

(11) Anderson, J. A. Infrared Study of CO Oxidation over Pt-Rh/Al2O3 Catalysts. *Journal of Catalysis* **1993**, *142* (1), 153–165. https://doi.org/10.1006/jcat.1993.1197.

(12) Chuang, S. S. C.; Srinivas, G.; Mukherjee, A. Infrared Studies of the Interactions of C2H4 and H2 with Rh+(CO)2 and CO Adsorbed on RhCl3/SiO2 and Rh(NO3)3/SiO2. *Journal of Catalysis* **1993**, *139* (2), 490–503. https://doi.org/10.1006/jcat.1993.1043.

(13) Kwon, Y.; Kim, T. Y.; Kwon, G.; Yi, J.; Lee, H. Selective Activation of Methane on Single-Atom Catalyst of Rhodium Dispersed on Zirconia for Direct Conversion. *Journal of the American Chemical Society* **2017**, *139* (48), 17694–17699. https://doi.org/10.1021/jacs.7b11010.

(14) Fang, C.-Y.; Zhang, S.; Hu, Y.; Vasiliu, M.; Perez-Aguilar, J. E.; Conley, E. T.; Dixon, D. A.; Chen, C.-Y.; Gates, B. C. Reversible Metal Aggregation and Redispersion Driven by the Catalytic Water Gas Shift Half-Reactions: Interconversion of Single-Site Rhodium Complexes and Tetrarhodium Clusters in Zeolite HY. *ACS Catalysis* **2019**, *9* (4), 3311–3321. https://doi.org/10.1021/acscatal.8b04798.

(15) Hülsey, M. J.; Zhang, B.; Ma, Z.; Asakura, H.; Do, D. A.; Chen, W.; Tanaka, T.; Zhang, P.; Wu, Z.; Yan, N. In Situ Spectroscopy-Guided Engineering of Rhodium Single-Atom Catalysts for CO Oxidation. *Nature Communications* **2019**, *10* (1), 1330. https://doi.org/10.1038/s41467-019-09188-9.

(16) Tang, Y.; Asokan, C.; Xu, M.; Graham, G. W.; Pan, X.; Christopher, P.; Li, J.; Sautet, P. Rh Single Atoms on TiO2 Dynamically Respond to Reaction Conditions by Adapting Their Site. *Nature Communications* **2019**, *10* (1), 4488. https://doi.org/10.1038/s41467-019-12461-6.

(17) Asokan, C.; Thang, H. V.; Pacchioni, G.; Christopher, P. Reductant Composition Influences the Coordination of Atomically Dispersed Rh on Anatase $TiO_2$. *Catalysis Science & Technology* **2020**, *10* (6), 1597–1601. https://doi.org/10.1039/D0CY00146E.

(18) Gu, F.; Qin, X.; Li, M.; Xu, Y.; Hong, S.; Ouyang, M.; Giannakakis, G.; Cao, S.; Peng, M.; Xie, J.; Wang, M.; Han, D.; Xiao, D.; Wang, X.; Wang, Z.; Ma, D. Selective Catalytic Oxidation of Methane to Methanol in Aqueous Medium over Copper Cations Promoted by Atomically Dispersed Rhodium on $TiO_2$. *Angewandte Chemie International Edition* **2022**, *61* (18). https://doi.org/10.1002/anie.202201540.

(19) Marino, S.; Wei, L.; Cortes-Reyes, M.; Cheng, Y.; Laing, P.; Cavataio, G.; Paolucci, C.; Epling, W. Rhodium Catalyst Structural Changes during, and Their Impacts on the Kinetics of, CO Oxidation. *JACS Au* **2023**, *3* (2), 459–467. https://doi.org/10.1021/jacsau.2c00595.

(20) Schroeder, E. K.; Finzel, J.; Christopher, P. Photolysis of Atomically Dispersed $Rh/Al_2O_3$ Catalysts: Controlling CO Coverage *in Situ* and Promoting Reaction Rates. *The Journal of Physical Chemistry C* **2022**, *126* (43), 18292–18305. https://doi.org/10.1021/acs.jpcc.2c04642.

(21) Zhang, J.; Asokan, C.; Zakem, G.; Christopher, P.; Medlin, J. W. Enhancing Sintering Resistance of Atomically Dispersed Catalysts in Reducing Environments with Organic Monolayers. *Green Energy & Environment* **2022**, *7* (6), 1263–1269. https://doi.org/10.1016/j.gee.2021.01.022.

(22) Brown, Tl. L. ; D., D. J. Intensities of CO Stretching Modes in the Infrared Spectra of Adsorbed CO and Metal Carbonyls. *Inorganic chemistry including bioinorganic chemistry* **1967**, *6* (5), 971. https://doi.org/10.1021/ic50051a026.

(23) Xu, R.; Jiang, Z.; Yang, Q.; Bloino, J.; Biczysko, M. Harmonic and Anharmonic Vibrational Computations for Biomolecular Building Blocks: Benchmarking DFT and Basis Sets by Theoretical and Experimental IR Spectrum of Glycine Conformers. *Journal of Computational Chemistry* **2024**, *45* (21), 1846–1869. https://doi.org/10.1002/jcc.27377.

(24) Rath, D.; Mikerásek, V.; Wang, C.; Eder, M.; Schmid, M.; Diebold, U.; Parkinson, G. S.; Pavelec, J. Infrared Reflection Absorption Spectroscopy Setup with Incidence Angle Selection for Surfaces of Non-Metals. *Review of Scientific Instruments* **2024**, *95* (6). https://doi.org/10.1063/5.0210860.

(25) Pavelec, J.; Hulva, J.; Halwidl, D.; Bliem, R.; Gamba, O.; Jakub, Z.; Brunbauer, F.; Schmid, M.; Diebold, U.; Parkinson, G. S. A Multi-Technique Study of $CO_2$ Adsorption on $Fe_3O_4$ Magnetite. *The Journal of Chemical Physics* **2017**, *146* (1). https://doi.org/10.1063/1.4973241.

(26) Halwidl, D. *Development of an Effusive Molecular Beam Apparatus*; Springer Fachmedien Wiesbaden: Wiesbaden, 2016. https://doi.org/10.1007/978-3-658-13536-2.

(27) Diebold, U. The Surface Science of Titanium Dioxide. *Surface Science Reports* **2003**, *48* (5–8), 53–229. https://doi.org/10.1016/S0167-5729(02)00100-0.

(28) Wendt, S.; Schaub, R.; Matthiesen, J.; Vestergaard, E. K.; Wahlström, E.; Rasmussen, M. D.; Thostrup, P.; Molina, L. M.; Lægsgaard, E.; Stensgaard, I.; Hammer, B.; Besenbacher, F. Oxygen Vacancies on $TiO_2(110)$ and Their Interaction with $H_2O$ and $O_2$: A Combined High-Resolution STM and DFT Study. *Surface Science* **2005**, *598* (1–3), 226–245. https://doi.org/10.1016/j.susc.2005.08.041.

(29) Giessibl, F. J. The qPlus Sensor, a Powerful Core for the Atomic Force Microscope. *Review of Scientific Instruments* **2019**, *90* (1). https://doi.org/10.1063/1.5052264.

(30) Choi, J. I. J.; Mayr-Schmölzer, W.; Mittendorfer, F.; Redinger, J.; Diebold, U.; Schmid, M. The Growth of Ultra-Thin Zirconia Films on $Pd_3Zr(0\,0\,0\,1)$. *Journal of Physics: Condensed Matter* **2014**, *26* (22), 225003. https://doi.org/10.1088/0953-8984/26/22/225003.

(31) Kresse, G.; Furthmüller, J. Efficiency of Ab-Initio Total Energy Calculations for Metals and Semiconductors Using a Plane-Wave Basis Set. *Computational Materials Science* **1996**, *6* (1), 15–50. https://doi.org/10.1016/0927-0256(96)00008-0.

(32) Blöchl, P. E. Projector Augmented-Wave Method. *Physical Review B* **1994**, *50* (24), 17953–17979. https://doi.org/10.1103/PhysRevB.50.17953.

(33) Kresse, G.; Joubert, D. From Ultrasoft Pseudopotentials to the Projector Augmented-Wave Method. *Physical Review B* **1999**, *59* (3), 1758–1775. https://doi.org/10.1103/PhysRevB.59.1758.

(34) Dion, M.; Rydberg, H.; Schröder, E.; Langreth, D. C.; Lundqvist, B. I. Van Der Waals Density Functional for General Geometries. *Physical Review Letters* **2004**, *92* (24), 246401. https://doi.org/10.1103/PhysRevLett.92.246401.

(35) Klimeš, J.; Bowler, D. R.; Michaelides, A. Chemical Accuracy for the van Der Waals Density Functional. *Journal of Physics: Condensed Matter* **2010**, *22* (2), 022201. https://doi.org/10.1088/0953-8984/22/2/022201.

(36) Klimeš, J.; Bowler, D. R.; Michaelides, A. Van Der Waals Density Functionals Applied to Solids. *Physical Review B* **2011**, *83* (19), 195131. https://doi.org/10.1103/PhysRevB.83.195131.

(37) Wang, Z.; Brock, C.; Matt, A.; Bevan, K. H. Implications of the $\mathrm{DFT}+U$ Method on Polaron Properties in Energy Materials. *Physical Review B* **2017**, *96* (12), 125150. https://doi.org/10.1103/PhysRevB.96.125150.

(38) Kowalski, P. M.; Meyer, B.; Marx, D. Composition, Structure, and Stability of the Rutile $TiO_2(110)$

</Mrow> </Math> Surface: Oxygen Depletion, Hydroxylation, Hydrogen Migration, and Water Adsorption. *Physical Review B* **2009**, *79* (11), 115410. https://doi.org/10.1103/PhysRevB.79.115410.
(39) Lizzit, S.; Baraldi, A.; Groso, A.; Reuter, K.; Ganduglia-Pirovano, M. V.; Stampfl, C.; Scheffler, M.; Stichler, M.; Keller, C.; Wurth, W.; Menzel, D. Surface Core-Level Shifts of Clean and Oxygen-Covered Ru(0001). *Physical Review B* **2001**, *63* (20), 205419. https://doi.org/10.1103/PhysRevB.63.205419.
(40) Köhler, L.; Kresse, G. Density Functional Study of CO on Rh(111). *Physical Review B* **2004**, *70* (16), 165405. https://doi.org/10.1103/PhysRevB.70.165405.
(41) Henkelman, G.; Arnaldsson, A.; Jónsson, H. A Fast and Robust Algorithm for Bader Decomposition of Charge Density. *Computational Materials Science* **2006**, *36* (3), 354–360. https://doi.org/10.1016/j.commatsci.2005.04.010.
(42) Krukau, A. V.; Vydrov, O. A.; Izmaylov, A. F.; Scuseria, G. E. Influence of the Exchange Screening Parameter on the Performance of Screened Hybrid Functionals. *The Journal of Chemical Physics* **2006**, *125* (22). https://doi.org/10.1063/1.2404663.
(43) Berkó, A.; Solymosi, F. Adsorption-Induced Structural Changes of Rh Supported by TiO2(110)-(1×2): An STM Study. *Journal of Catalysis* **1999**, *183* (1), 91–101. https://doi.org/10.1006/jcat.1998.2368.
(44) Evans J.; Hayden B.; Mosselmans F.; Murray A. Rhodium Geminal Dicarbonyl on TiO2 (110). *Journal of the American Chemical Society JACS* **1992**, *114* (17), 6912. https://doi.org/10.1021/ja00043a044.
(45) Kondarides, D. I.; Zhang, Z.; Verykios, X. E. Chlorine-Induced Alterations in Oxidation State and CO Chemisorptive Properties of CeO2-Supported Rh Catalysts. *Journal of Catalysis* **1998**, *176* (2), 536–544. https://doi.org/10.1006/jcat.1998.2064.
(46) Sombut, P.; Puntscher, L.; Atzmueller, M.; Jakub, Z.; Reticcioli, M.; Meier, M.; Parkinson, G. S.; Franchini, C. Role of Polarons in Single-Atom Catalysts: Case Study of Me1 [Au1, Pt1, and Rh1] on TiO2(110). *Topics in Catalysis* **2022**, *65* (17–18), 1620–1630. https://doi.org/10.1007/s11244-022-01651-0.
(47) Petrik, N. G.; Kimmel, G. A. Adsorption Geometry of CO versus Coverage on TiO$_2$ (110) from s- and p-Polarized Infrared Spectroscopy. *The Journal of Physical Chemistry Letters* **2012**, *3* (23), 3425–3430. https://doi.org/10.1021/jz301413v.
(48) Rice, C. A.; Worley, S. D.; Curtis, C. W.; Guin, J. A.; Tarrer, A. R. The Oxidation State of Dispersed Rh on Al2O3. *The Journal of Chemical Physics* **1981**, *74* (11), 6487–6497. https://doi.org/10.1063/1.440987.
(49) Vayssilov, G. N.; Rösch, N. A New Interpretation of the IR Bands of Supported Rh(I) Monocarbonyl Complexes. *Journal of the American Chemical Society* **2002**, *124* (14), 3783–3786. https://doi.org/10.1021/ja011688y.
(50) Hayden, B. E.; King, A.; Newton, M. A. The Alignment of a Surface Species Determined by FT-RAIRS: Rhodium Gem-Dicarbonyl on TiO2(110). *Chemical Physics Letters* **1997**, *269* (5), 485–488. https://doi.org/10.1016/S0009-2614(97)00305-9.
(51) Wang, Y.; Wöll, C. IR Spectroscopic Investigations of Chemical and Photochemical Reactions on Metal Oxides: Bridging the Materials Gap. *Chemical Society Reviews* **2017**, *46* (7), 1875–1932. https://doi.org/10.1039/C6CS00914J.
(52) Krenn, G.; Bako, I.; Schennach, R. CO Adsorption and CO and O Coadsorption on Rh(111) Studied by Reflection Absorption Infrared Spectroscopy and Density Functional Theory. *The Journal of Chemical Physics* **2006**, *124* (14). https://doi.org/10.1063/1.2184308.

(53) Parkinson, G. S.; Dohnálek, Z.; Smith, R. S.; Kay, B. D. Reactivity of $C_2Cl_6$ and $C_2Cl_4$ Multilayers with $Fe^0$ Atoms over FeO(111). *The Journal of Physical Chemistry C* **2009**, *113* (23), 10233–10241. https://doi.org/10.1021/jp901040f.

(54) Parkinson, G. S.; Dohnálek, Z.; Smith, R. S.; Kay, B. D. Reactivity of $Fe^0$ Atoms with Mixed $CCl_4$ and $D_2O$ Films over FeO(111). *The Journal of Physical Chemistry C* **2010**, *114* (40), 17136–17141. https://doi.org/10.1021/jp103896k.

(55) Reticcioli, M.; Setvin, M.; Schmid, M.; Diebold, U.; Franchini, C. Formation and Dynamics of Small Polarons on the Rutile $TiO_2$ (110) Surface. *Physical Review B* **2018**, *98* (4), 045306. https://doi.org/10.1103/PhysRevB.98.045306.

(56) Anderson, G. K.; Cross, R. J. Isomerisation Mechanisms of Square-Planar Complexes. *Chemical Society Reviews* **1980**, *9* (2), 185–215. https://doi.org/10.1039/CS9800900185.

(57) Diebold, U.; Anderson, J. F.; Ng, K.-O.; Vanderbilt, D. Evidence for the Tunneling Site on Transition-Metal Oxides: $TiO_2$ (110). *Physical Review Letters* **1996**, *77* (7), 1322–1325. https://doi.org/10.1103/PhysRevLett.77.1322.

(58) Hayden, B. E.; King, A.; Newton, M. A. The Reaction of Hydrogen with TiO2(110) Supported Rhodium Gem-Dicarbonyl. *Surface Science* **1998**, *397* (1–3), 306–313. https://doi.org/10.1016/S0039-6028(97)00749-8.

(59) Frederick, B. G.; Apai, G.; Rhodin, T. N. An X-Ray Photoelectron Spectroscopy Study of Rhodium Carbonyls Adsorbed on Planar Aluminas: Formation of Geminal Dicarbonyl Species. *Journal of the American Chemical Society* **1987**, *109* (16), 4797–4803. https://doi.org/10.1021/ja00250a007.

(60) Hulva, J.; Meier, M.; Bliem, R.; Jakub, Z.; Kraushofer, F.; Schmid, M.; Diebold, U.; Franchini, C.; Parkinson, G. S. Unraveling CO Adsorption on Model Single-Atom Catalysts. *Science* **2021**, *371* (6527), 375–379. https://doi.org/10.1126/science.abe5757.

(61) Babucci, M.; Guntida, A.; Gates, B. C. Atomically Dispersed Metals on Well-Defined Supports Including Zeolites and Metal–Organic Frameworks: Structure, Bonding, Reactivity, and Catalysis. *Chemical Reviews* **2020**, *120* (21), 11956–11985. https://doi.org/10.1021/acs.chemrev.0c00864.

(62) Evans, J.; Hayden, B. E.; Mosselmans, J. F. W.; Murray, A. J. Model Catalyst Studies of Titania Supported Rhodium. In *Elementary Reaction Steps in Heterogeneous Catalysis*; Joyner, R. W., van Santen, R. A., Eds.; Springer Netherlands: Dordrecht, 1993; pp 179–195. https://doi.org/10.1007/978-94-011-1693-0_11.

(63) Lamb, H. H.; Gates, B. C.; Knözinger, H. Molecular Organometallic Chemistry on Surfaces: Reactivity of Metal Carbonyls on Metal Oxides. *Angewandte Chemie International Edition in English* **1988**, *27* (9), 1127–1144. https://doi.org/10.1002/anie.198811271.

(64) DeLouise, L. A.; White, E. J.; Winograd, N. CHaracterization of CO Binding Sites on Rh{111} and Rh{331} Surfaces by XPS and LEED: Comparison to EELS Results. *Surface Science* **1984**, *147* (1), 252–262. https://doi.org/10.1016/0039-6028(84)90179-1.

(65) Cavanagh, R. R.; Yates, J. T. Site Distribution Studies of Rh Supported on Al2O3—An Infrared Study of Chemisorbed CO. *The Journal of Chemical Physics* **1981**, *74* (7), 4150–4155. https://doi.org/10.1063/1.441544.

(66) Ryan, P.; Jakub, Z.; Balajka, J.; Hulva, J.; Meier, M.; Küchle, J.; Blowey, P. J.; Thakur, P. K.; Franchini, C.; Payne, D. Direct Measurement of Ni Incorporation into $Fe_3O_4$ (001). *Physical Chemistry Chemical Physics* **2018**, *20* (24), 16469–16476.

(67) Isaacs, M. A.; Graf, A.; Morgan, D. J. XPS Insight Note: Multiplet Splitting in X-Ray Photoelectron Spectra. *Surface and Interface Analysis* **2025**, *57* (4), 285–290. https://doi.org/10.1002/sia.7383.

(68) Primet, M.; Vedrine, J. C.; Naccache, C. Formation of Rhodium Carbonyl Complexes in Zeolite. *Journal of Molecular Catalysis* **1978**, *4* (6), 411–421. https://doi.org/10.1016/0304-5102(78)80011-X.

(69) Liu, L.; Corma, A. Metal Catalysts for Heterogeneous Catalysis: From Single Atoms to Nanoclusters and Nanoparticles. *Chemical Reviews* **2018**, *118* (10), 4981–5079. https://doi.org/10.1021/acs.chemrev.7b00776.